\documentclass[a4paper,11pt]{scrartcl}

\usepackage{mathtools}

\usepackage[british]{babel}
\usepackage[utf8]{inputenc}
\usepackage{amsthm}
\usepackage{amssymb}
\usepackage{amsmath}
\usepackage{nicefrac}
\usepackage{float}
\usepackage{setspace}
\usepackage{url}
\usepackage{bm}
\usepackage{makecell}
\usepackage{caption}
\usepackage{enumerate}
\usepackage{listings}
\usepackage{tabu}
\usepackage{color}
\usepackage{cprotect}
\usepackage{graphicx}
\usepackage{subfig}
\usepackage{makeidx}
\usepackage[ruled]{algorithm2e}
\usepackage{lipsum} 
\usepackage{subfiles} 
\usepackage{comment}
\usepackage{afterpage}

\title{Crack Modeling via Minimum-Weight Surfaces in 3d Voronoi Diagrams}
\author{Christian Jung$^{1}$ \and Claudia Redenbach$^1$}

\date{
	{\small $^1$Technische Universität Kaiserslautern \\  Gottlieb-Daimler-Straße 47, 67663 Kaiserslautern, Germany}
}

\begin{document}

\maketitle

\textbf{Abstract:} Shortest paths play an important role in mathematical modeling and image processing. Usually, shortest path problems are formulated on planar graphs that consist of vertices and weighted arcs. In this context, one is interested in finding a path of minimum weight from a start vertex to an end vertex. The concept of minimum-weight surfaces extends shortest paths to 3d. The minimum-weight surface problem is formulated on a cellular complex with weighted facets. A cycle on the arcs of the complex serves as input and one is interested in finding a surface of minimum weight bounded by that cycle. In practice, minimum-weight surfaces can be used to segment 3d images. Vice versa, it is possible to use them as a modeling tool for geometric structures such as cracks. In this work, we present an approach for using minimum-weight surfaces in bounded Voronoi diagrams to generate synthetic 3d images of cracks.\\


\section{Introduction}
The shortest path problem (SPP) is formulated on a planar (directed) graph consisting of vertices and weighted arcs. Given a start and an end vertex, one is interested in finding a path of minimal weight connecting these vertices. Shortest paths represent a useful tool in 2d image processing in several contexts including image quilting \cite{quilting} and image segmentation. In particular, shortest paths have been used to segment linear, 1d structures such as cracks in 2d images \cite{minPaths1}.

Minimum-weight surfaces were introduced in \cite{sullivan} and \cite{grady} as an extension of shortest paths to 3d. The minimum-weight surface problem (MSP) is formulated on a cellular complex consisting of vertices, arcs, weighted facets and cells. Here, the goal is to find a set of facets of minimal weight that is bounded by an input cycle. The voxel lattice in a 3d image can be interpreted as a cellular complex. In this context, minimum-weight surfaces have been used to segment flat structures in medical 3d images \cite{grady}.

The segmentation of cracks in images of concrete is a broad field of research. A large variety of segmentation methods have been studied for 2d and 3d images. 
Comprehensive overviews and comparison studies in 2d and 3d can be found in \cite{2d-review} and \cite{ourPaper}, respectively. 

In image segmentation, ground truths are necessary for an objective output evaluation. Furthermore, they are prerequisite for training machine learning models such as convolutional neural networks \cite{3dunet} and random forests \cite{ranfor}. In 2d, annotated data is abundantly available, for example SDNET2018 \cite{sdnet2018}. For 3d images, manual annotation is not feasible due to the large amount of data. Therefore, 3d ground truth images are scarce and one has to rely on artificial data. Previously, the generation of synthetic cracks has been realized via fractional Brownian surfaces \cite{Addison_2000, ourPaper}. However, the Hurst index - a measure of roughness - is the only parameter that can be used to control the shape of the surface. Therefore, this model is rather limited.

Voronoi diagrams have been used previously for simulating crack propagation via finite element methods \cite{sim1,sim1-5}. In this work, we propose a novel method to synthesize crack structures in 3d images using Voronoi diagrams generated from random point processes. The method includes two aspects:

First, we use minimum-weight surfaces as a tool to model the macrostructure of cracks. Bounded 3d Voronoi diagrams serve as the underlying cell complexes. This leaves several degrees of freedom such as the choice of the generating point processes, bounding cycles and the facet weights. 

Second, the computed surfaces are discretized to 3d binary images. The rough microstructure of cracks is modeled via a second Voronoi diagram on a finer scale. Then, the cracks are embedded into patches of real computed tomography (CT) images of concrete.

This paper is structured as follows. In Section 2, we outline the concept of shortest paths. Their extension to minimum-weight surfaces is described in Section 3. Section 4 comprises our crack modeling pipeline using Voronoi diagrams. It includes a description of our approach for macro- and microstructure modeling. Section 5 serves as conclusion and outlook to possible future research.

\section{Shortest paths}\label{sec:sp}

Let $G=(V,A)$ be a directed graph consisting of a set of vertices $V$ and directed arcs $A\subseteq V\times V$. We use the notation $\alpha(a)$ for the start vertex and $\omega(a)$ for the end vertex of an arc $a\in A$.


Furthermore, let $c:A\rightarrow\mathbb{R}_{>0}$ be a function assigning a non-negative weight to every arc in $A$. A path in $G$ is a finite sequence of vertices and arcs, $P=(v_0,a_0,v_1,a_1,\ldots,a_{k-1},v_k)$, $k\geq 0$, with $v_i\in V$ and $a_i\in A$ with $\alpha(a_i)=v_i$ and $\omega(a_i)=v_{i+1}$ for $i=0,\ldots,k-1$. It is called a cycle (or closed contour) if no arc and no vertex is included more than once except for $v_0 = v_k$. The weight of a path $P$ is given as $c(P) = \sum_{i=0}^{k-1} c(a_{i})$. 

Given two vertices $s,t\in V$, the SPP is looking for a path of minimum weight from $s$ to $t$. Many algorithms for solving the SPP exist, for example Dijkstra's \cite{dijkstra} or Bellman's and Ford's algorithm \cite{bellman}.

SPPs can also be formulated as binary integer programs \cite{spp1, spp2}. Note that this approach is less efficient than the ones described in \cite{dijkstra} or \cite{bellman}. However, it gives a good intuition for an analogue approach to compute minimum-weight surfaces which we describe in Section \ref{sec:ms}.

Let $s$ be the start- and $t$ the end vertex of the path and let $x$ be a vector of binary variables $x_{i} \in \{0,1\}$ assigned to each arc $a_{i}$. The SPP can then be formulated as
\begin{alignat}{3}
& \text{minimize}   & \quad\quad & \sum_{i: \, a_{i}\in A} c(a_{i})x_{i} & &\label{eq1} \\
& \text{subject to} &       & Bx=p & & \label{eq2} \\
&                   &       & x_{i} \in \{0,1\}.  & & \label{eq3}
\end{alignat}

The vertex-arc incidence matrix $B$ and the vector $p$ in constraints (\ref{eq2}) are given as

\[
B_{j,i} = \begin{dcases}
\begin{array}{ll}
1 & \textrm{if } v_j=\alpha(a_i), \\
-1 &  \textrm{if } v_j=\omega(a_i), \\
0 & \, \textrm{else}
\end{array}
\end{dcases}
\
\textrm{ and }
\
p_{j} = \begin{dcases}
\begin{array}{ll}
1 & \textrm{if } v_j=s, \\
-1 &  \textrm{if } v_j=t, \\
0 & \, \textrm{else}.
\end{array}
\end{dcases}
\]

The constraints (\ref{eq2}) are flow-conservation constraints. They ensure that every vertex that is part of the path is incident to exactly one incoming and one outgoing arc, except for the start- and end vertex. This results in the fact that any feasible solution must contain a path from $s$ to $t$ and, by the assumption that all costs are strictly positive, this ensures that the optimal solution is in fact a path without repeated vertices. The variables $x_{i}$ are binary by constraint (\ref{eq3}) and indicate whether arc $a_{i}$ belongs to the path ($x_{i}=1$) or not ($x_{i}=0$). Finally, the objective in (\ref{eq1}) is to minimize the weight over all paths from $s$ to $t$.


Note that, in case of negative arc costs, the problem of finding a shortest path without repeated vertices becomes NP-hard (which can be seen by a reduction from the Hamiltonian Path Problem \cite{NP}). Thus, additional constraints become necessary and, hence, we assume that all arc costs are strictly positive.

\section{Minimum-weight surfaces}\label{sec:ms}

Minimum-weight surfaces have been presented in \cite{sullivan} and \cite{grady} as an extension of shortest paths to 3d. 

For our purposes, let $K=(V,A,F,C)$ be a cellular complex consisting of a set of vertices $V$, directed arcs $A\subseteq V\times V$, facets $F\subseteq A\times\ldots \times A$ and cells $C\subseteq F\times \ldots \times F$. 

Further, let $w: F\rightarrow \mathbb{R}_{>0}$ be a function assigning a non-negative weight to every facet in $F$.

Note that $(V,A)$ defines a (directed) graph. Given a cycle $H$ on $(V,A)$, the MSP is looking for a connected set of facets in $K$ of mimimum weight that is bounded by $H$. 

To this end, arc directions may be assigned arbitrarily. Every facet is considered twice, once in clockwise and once in counterclockwise orientation. The orientation of $H$ must be chosen to be either clockwise or counterclockwise. If the direction of arc $a$ coincides with the direction of its counterpart in an incident facet $f$ (or cycle $H$), we call $a$ and $f$ (or $a$ and $H$) coherent. If it does not coincide, we call them anti-coherent.

The MSP can be formulated as a binary integer program analogously to the one in Section \ref{sec:sp}. It is given as
\begin{alignat}{3}
& \text{minimize}   & \quad\quad & \sum_{i: \, f_{i}\in F} w(f_{i})y_{i} & &\label{eq1b} \\
& \text{subject to} &       & Dy=q & & \label{eq2b} \\
&                   &       & y_{i} \in \{0,1\}.  & & \label{eq3b}
\end{alignat}


The arc-facet incidence matrix $D$ and the vector $q$ in constraints (\ref{eq2b}) are given as

\[
D_{j,i} = \begin{dcases}
\begin{array}{ll}
1 & \textrm{if } a_j \textrm{ and } f_i \textrm{ are incident}  \\
& \textrm{and coherent,}\\
-1 & \textrm{if } a_j \textrm{ and } f_i \textrm{ are incident}\\
& \textrm{and anti-coherent,}\\
0 & \, \textrm{else}
\end{array}
\end{dcases}
\]

and 

\[
q_{j} = \begin{dcases}
\begin{array}{ll}
1 & \textrm{if } a_j \textrm{ is part of } H \\
& \textrm{and coherent to $H$} \\
-1 & \textrm{if } a_j \textrm{ is part of } H \\
& \textrm{and anti-coherent to $H$} \\
0 & \, \textrm{else.}
\end{array}
\end{dcases}
\]


Similarly to the flow-conservation constraints in (\ref{eq2}), constraints (\ref{eq2b}) ensure that we obtain a connected set of facets that is bounded by $H$. The variables $y_{i}$ are binary by constraint (\ref{eq3b}) and indicate whether facet $f_{i}$ belongs to the surface ($y_{i}=1$) or not ($y_{i}=0$). Finally, the objective in (\ref{eq1b}) is to minimize the weight over all possible surfaces that are bounded by $H$.

We always assume positive costs to ensure that the output is indeed a connected surface.


\section{Crack modeling}\label{sec:crackModeling}

Minimum-weight surfaces have been used for 3d image segmentation with the 3d grid being the underlying cell complex \cite{grady}. Reversing this approach, minimum-weight surfaces can also be used to model surface-like, connected structures such as cracks on a macroscopic level.

Here, we focus on minimum-weight surfaces in bounded 3d Voronoi diagrams. Our goal is to develop an approach for modeling 2d crack structures to generate semi-synthetic 3d images of cracked concrete. Using this approach in combination with various point process models allows us to control the geometry of the resulting crack structure.

\begin{figure}[h]
	\centering
	\includegraphics[width=0.7\textwidth]{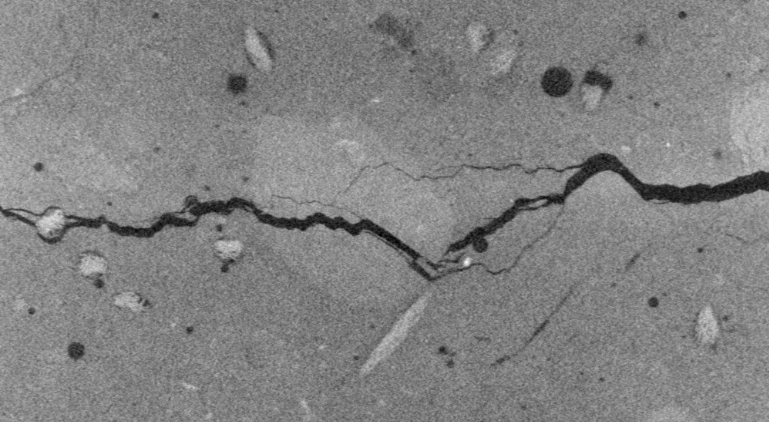}
	\caption{2d slice of a 3d CT image of cracked concrete of size $1000\times 550\times 880$ voxels with a voxel edge length of $22.7\mu$m. The image stems from a normal-strength concrete sample that was exposed to a tensile test. Sample: Department of Civil Engineering, University of Kaiserslautern, Imaging: Fraunhofer ITWM, Kaiserslautern} \label{fig:realCrack}
\end{figure}

The most striking geometric characteristics of cracks in 3d CT images of concrete have been identified and discussed in \cite{iCT}. In particular, we observe that 1. crack widths are varying and cracks may appear on multiple scales, 2. crack surfaces are not totally smooth but rather rough due to the granularity of the concrete's cement matrix, 3. when propagating through concrete, cracks may branch. An example is given in Fig. \ref{fig:realCrack}. These observations are used in the discretization procedure for modeling the cracks' microstructure.

\subsection{Voronoi diagrams}





Voronoi diagrams are defined on a metric 
space $(S,d)$ with Euclidean distance function $d$. Given a set of points $R=\{p_1, \ldots ,p_n\} \subset S$ with $2 \leq n < \infty$, the Voronoi diagram generated by $R$ is given as 
$W=\{C_1,\ldots,C_n\}$ with
\begin{equation*}\label{tessdist}
C_i=\{x\in S \mid d(x,p_i) \leq d(x,p_j) \ \forall\ p_j\in R\}.
\end{equation*}
The $C_i$ are called the cells of the Voronoi diagram. 


Note that, if $S$ is not bounded, the Voronoi diagram contains cells of infinite size. In practice, it is often convenient to only consider bounded cells. Therefore, we restrict our attention to the bounded Voronoi diagram given by $W\cap Q = \{C_1\cap Q,\ldots,C_n\cap Q\}$ for some bounded region $Q\subset S$. Note that this operation yields additional vertices, arcs and facets on the boundary of $Q$ that belong to the bounded Voronoi diagram.

\subsection{Minimum-weight Voronoi surfaces}

Bounded Voronoi diagrams in 3d can be considered as a cellular complex.
Therefore, given a cycle on the arcs of the cell complex induced by a 3d Voronoi diagram and weighted facets, we are able to compute a minimum-weight surface by solving the optimization problem given in Section \ref{sec:ms}. As a result, we obtain a connected set of facets that we call a minimum-weight Voronoi surface.

\subsection{Crack generation}\label{crackgen}

%

We propose the following method to simulate crack structures via minimum-weight Voronoi surfaces. 

\begin{enumerate}
	\item Define a cuboid $Q=[0,d_1]\times[0,d_2]\times[0,d_3]\subset\mathbb{R}^3$.
	\item Compute a random point pattern $R\subset Q$ as a realization of some point process model.
	\item Compute the Voronoi diagram generated by $R$, bounded by $Q$.
	\item Define functions $c,w$ assigning a non-negative weight to each of the arcs and facets, respectively.
	\item Choose a vertex on each of the four vertical edges of $Q$. Denote them by $u_1,u_2,u_3,u_4$. 
	Compute shortest paths from $u_1$ to $u_2$, $u_2$ to $u_3$, $u_3$ to $u_4$ and $u_4$ to $u_1$, via Dijkstra's algorithm, only using arcs that lie on the boundary of $Q$. Denote the paths by $P_1,P_2,P_3,P_4$. Then, $H=\cup_{i=1}^4 P_i$ is a cycle on the boundary of $Q$. 
	\item Compute a minimum-weight surface bounded by $P$ by solving the integer program from Section \ref{sec:ms}.
\end{enumerate}

\begin{figure*}[ht]
	\centering
	\subfloat{\includegraphics[width=0.33\textwidth]{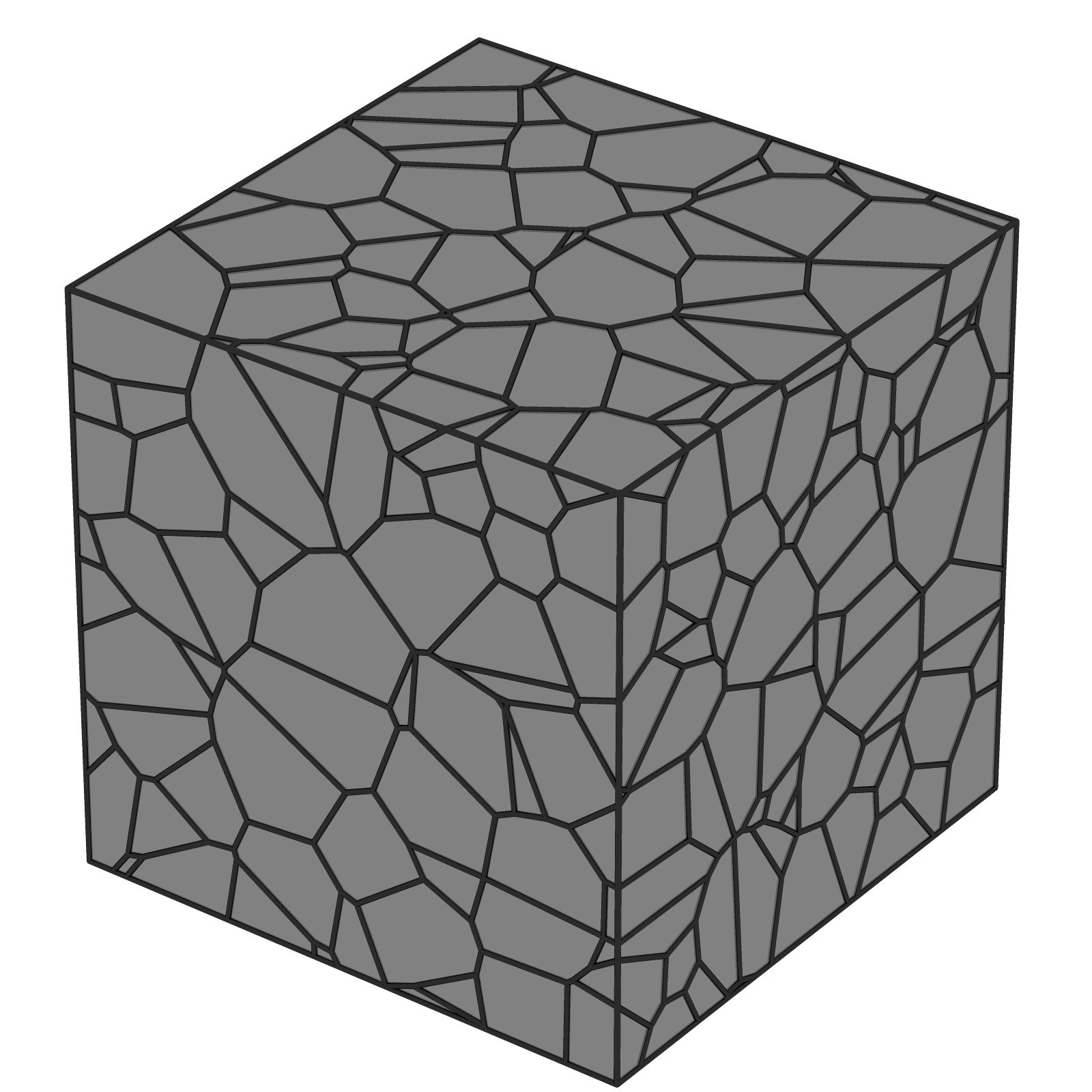}}\hfil
	\subfloat{\includegraphics[width=0.33\textwidth]{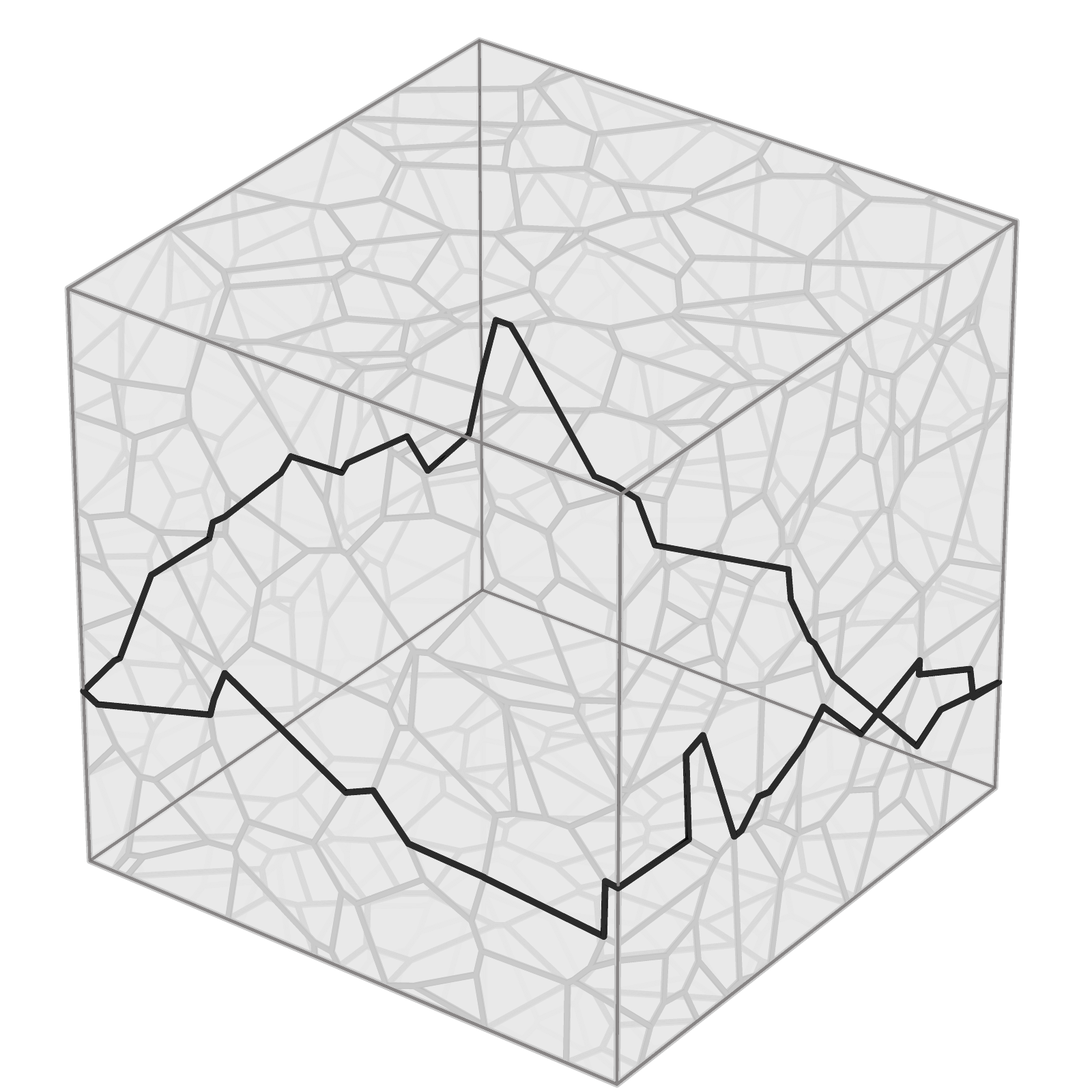}}\hfil
	\subfloat{\includegraphics[width=0.33\textwidth]{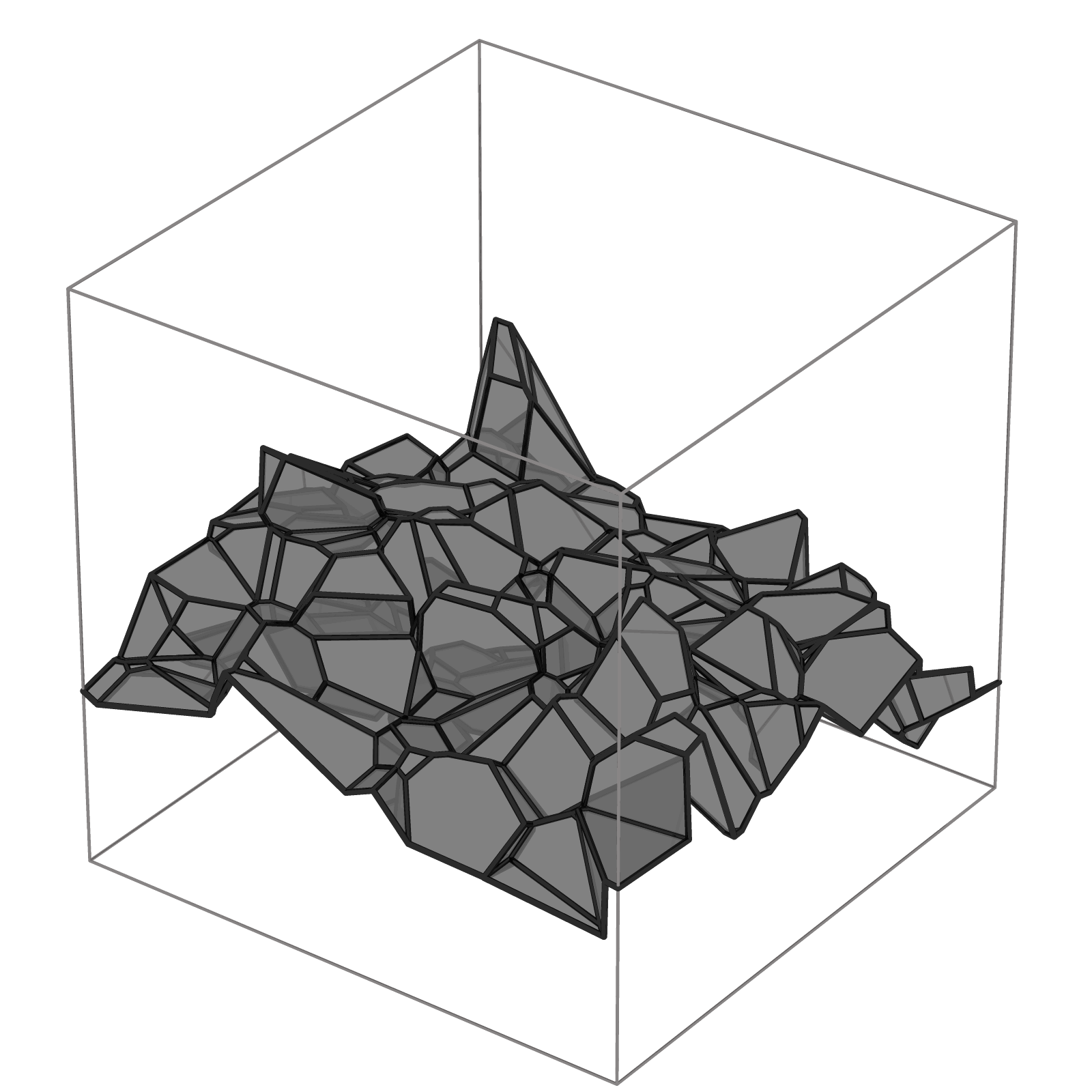}}\hfil
	\caption{Minimum-weight surface generation. Left: Poisson-Voronoi diagram bounded by a cube, middle: cycle on the boundary of the Voronoi diagram, right: minimum-weight surface.}\label{fig:minsurf}
\end{figure*}

The concept is illustrated in Fig. \ref{fig:minsurf}. It shows a minimum-weight surface in a Voronoi diagram bounded by $Q=[0,1]\times[0,1]\times[0,1]$. Its generators are a realization of a Poisson point process in $Q$ of intensity $500$. Arcs and facets are both assigned unit weights.

The approach above leaves us several degrees of freedom. Facet shape and variability can be controlled by choice of the underlying point process model, while the intensity of the generator process influences the mean facet size. Additionally, the size of the bounding cuboid, the input cycle and the weighting functions $c$ and $w$ can be varied.

\begin{figure*}
	\centering
	\subfloat{\includegraphics[width=0.33\textwidth]{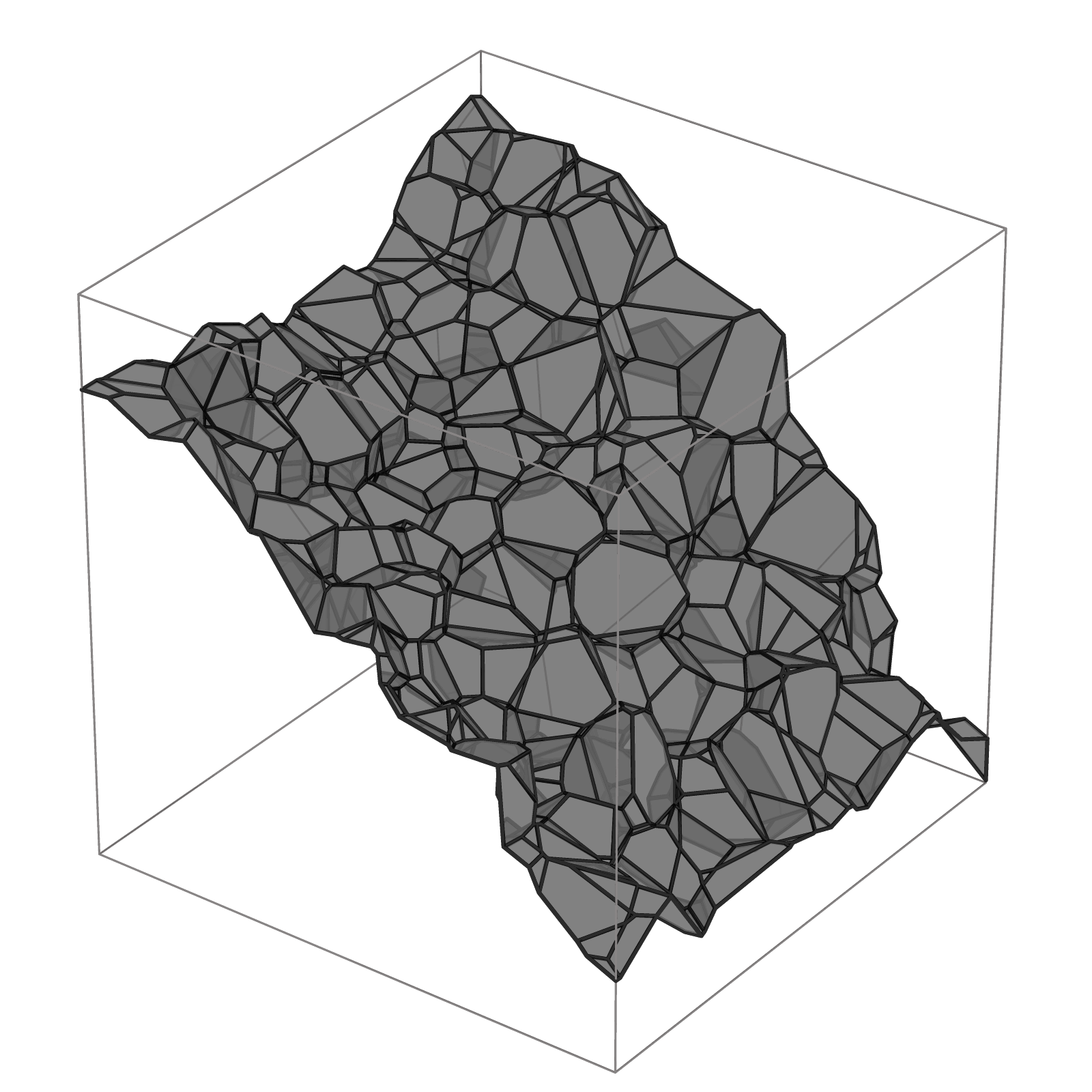}}\hfil
	\subfloat{\includegraphics[width=0.33\textwidth]{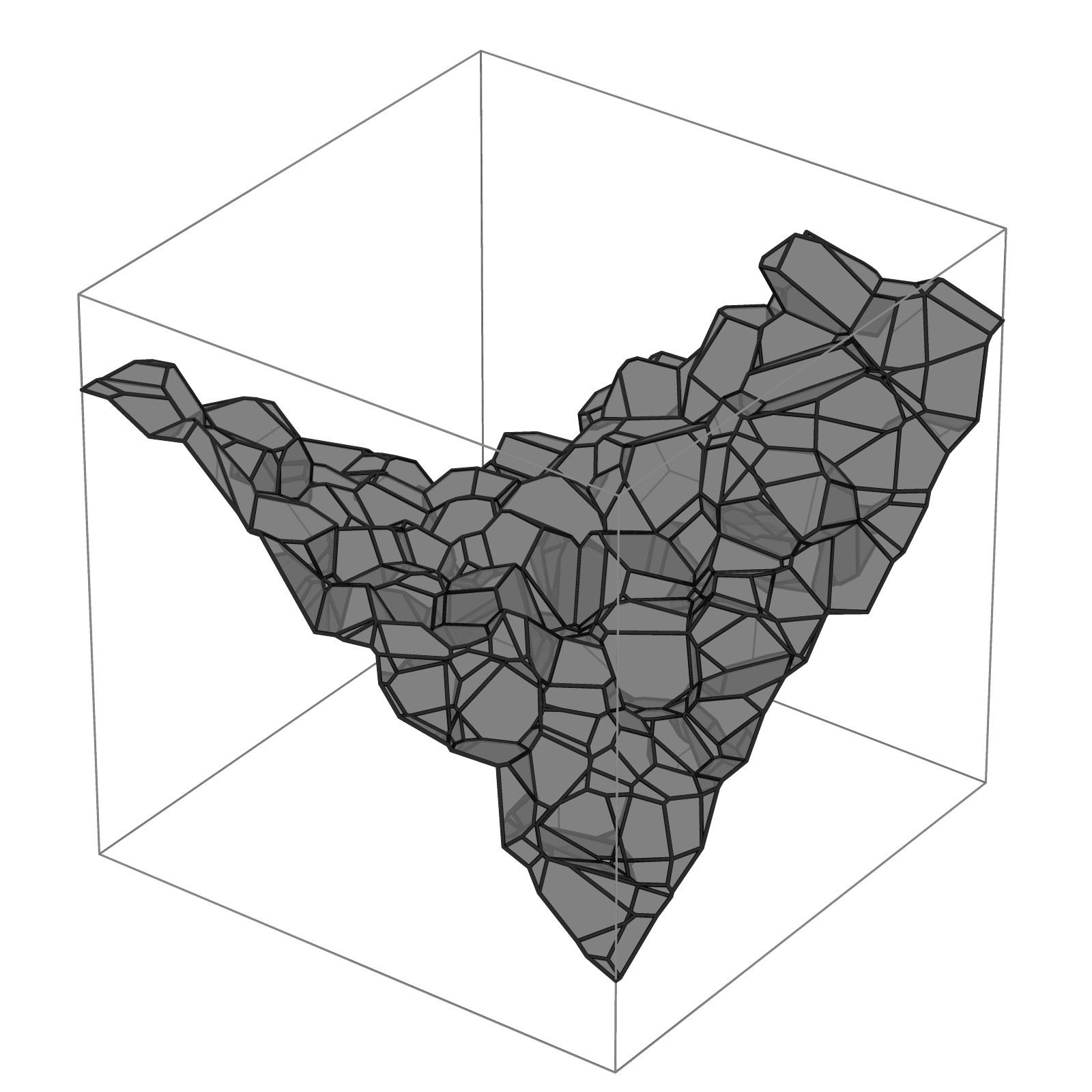}}\hfil
	\subfloat{\includegraphics[width=0.33\textwidth]{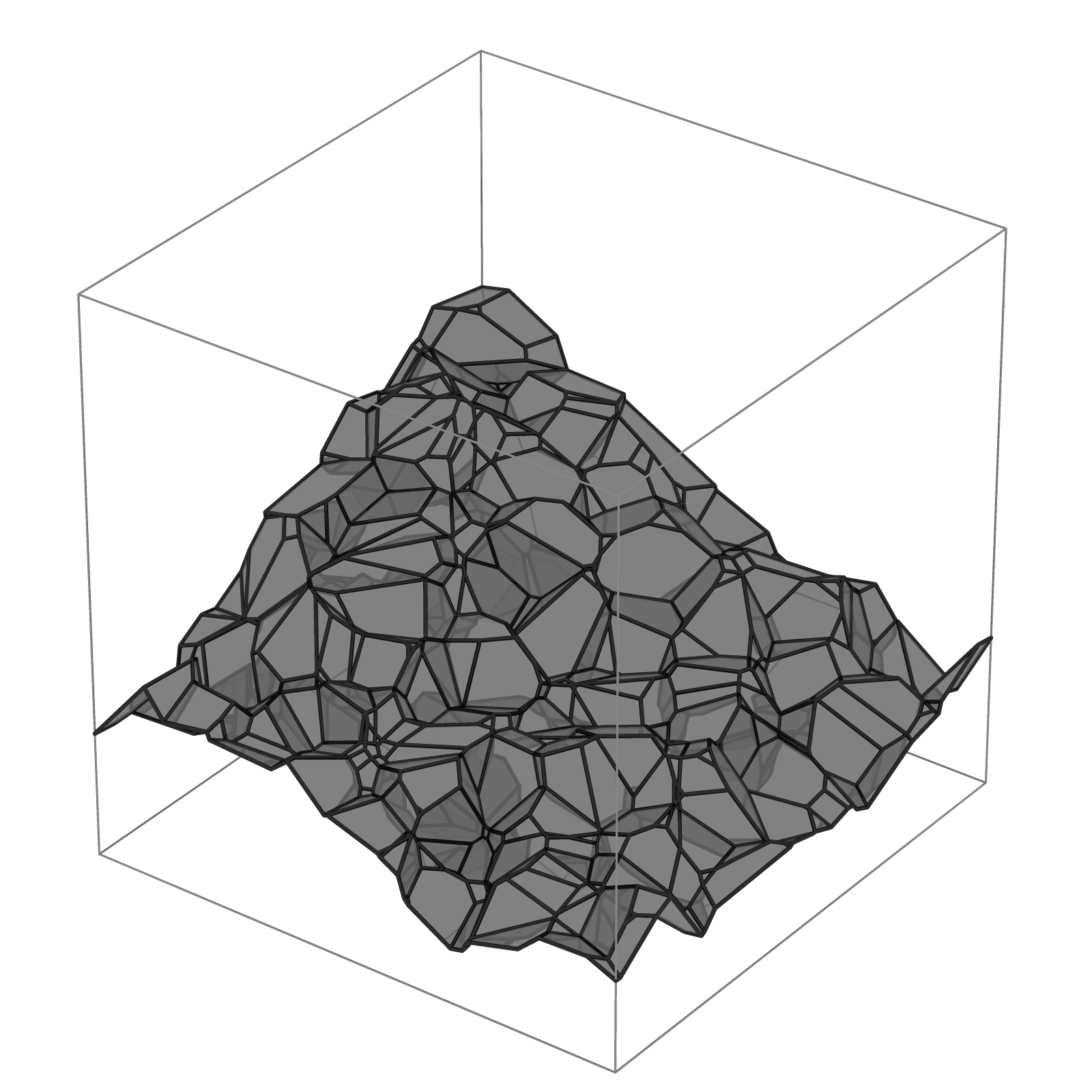}}\hfil
	\caption{Examples of minimum-weight surfaces in the same Voronoi diagram but with different choices of input cycles.}\label{fig:endvert}
\end{figure*}

Minimum-weight surfaces for the same realization of a Poisson point process but different choices of input cycles are given in Fig. \ref{fig:endvert}. Note that the input cycle is not restricted to lie on the boundary of a cube but may be chosen arbitrarily. 

\begin{figure*}
	\captionsetup[subfigure]{labelformat=empty}
	\centering
	\subfloat[$\lambda=50$]{\includegraphics[width=0.24\textwidth]{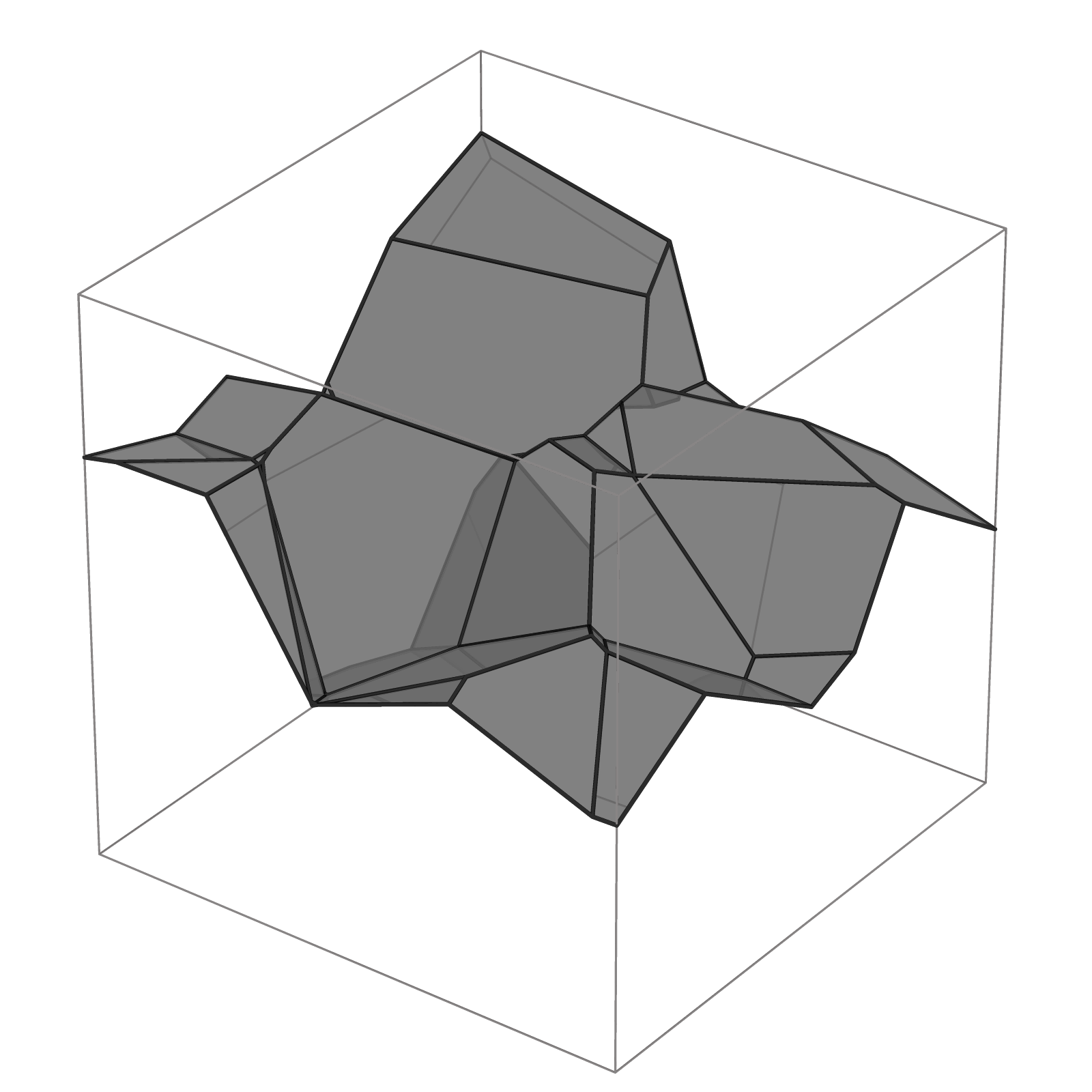}}\hfil
	\subfloat[$\lambda=2,\mu=50,r=0.1$]{\includegraphics[width=0.24\textwidth]{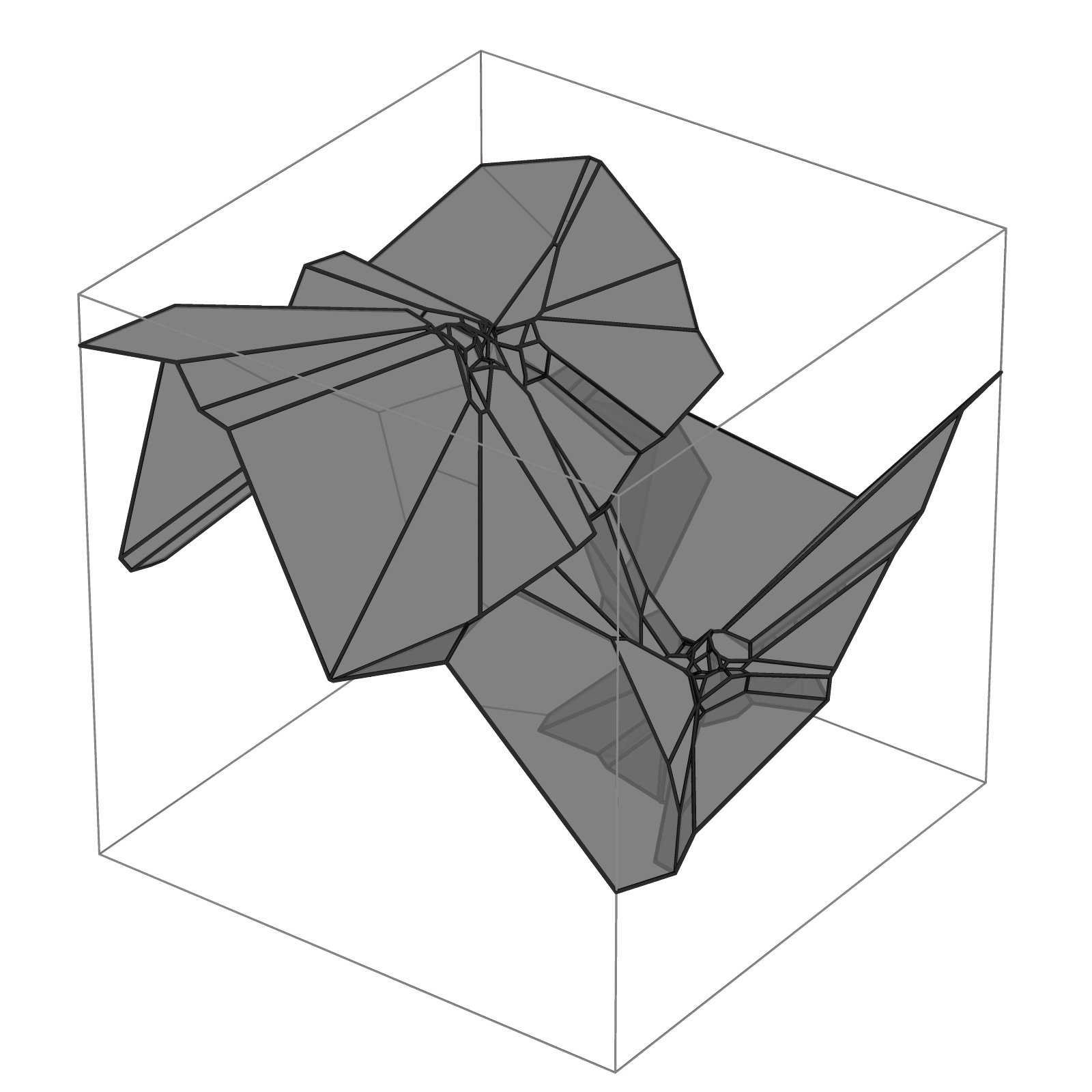}}\hfil
	\subfloat[$\lambda=50$]{\includegraphics[width=0.24\textwidth]{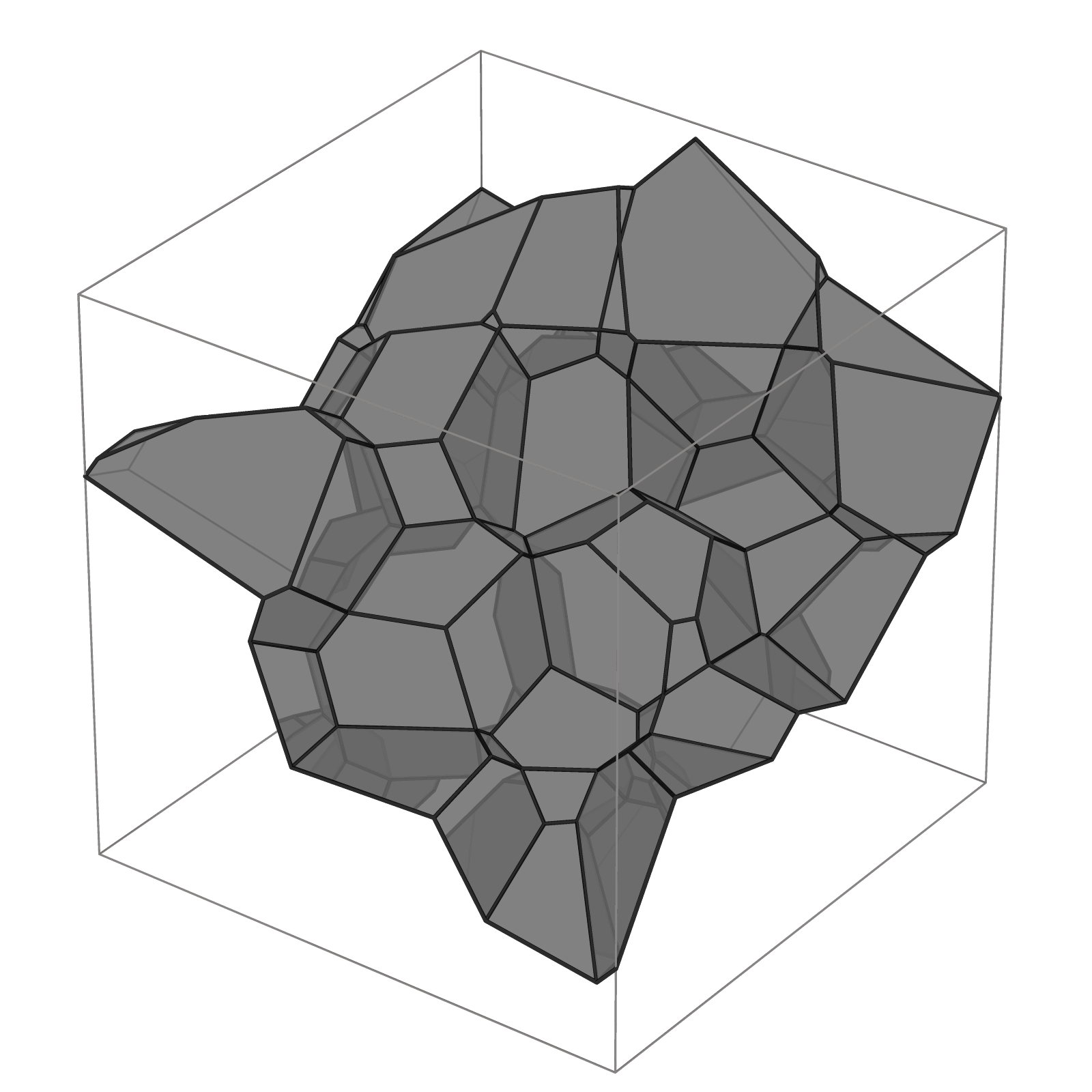}}\hfil
	
	\subfloat[$\lambda=100$]{\includegraphics[width=0.24\textwidth]{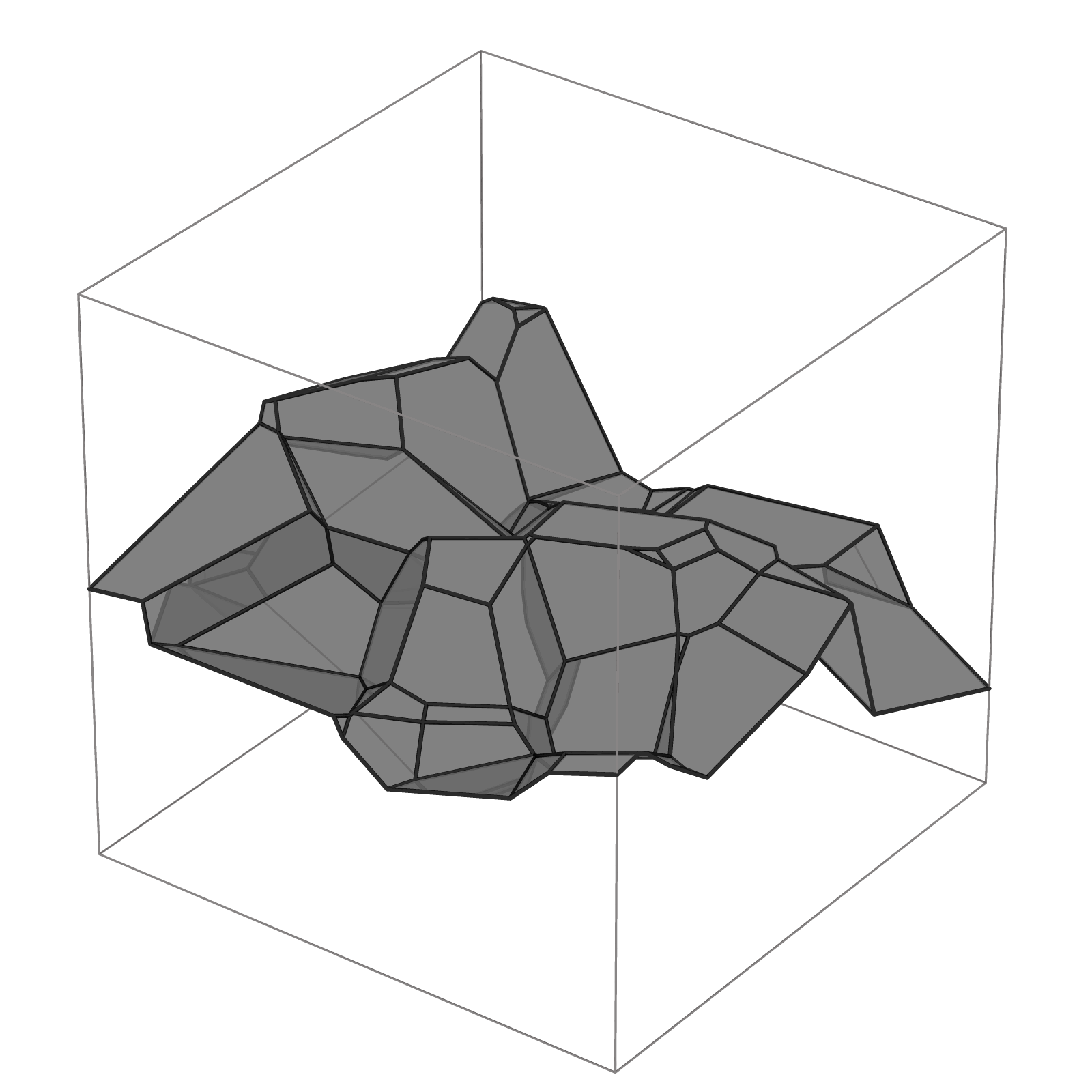}}\hfil
	\subfloat[$\lambda=5,\mu=100,r=0.1$]{\includegraphics[width=0.24\textwidth]{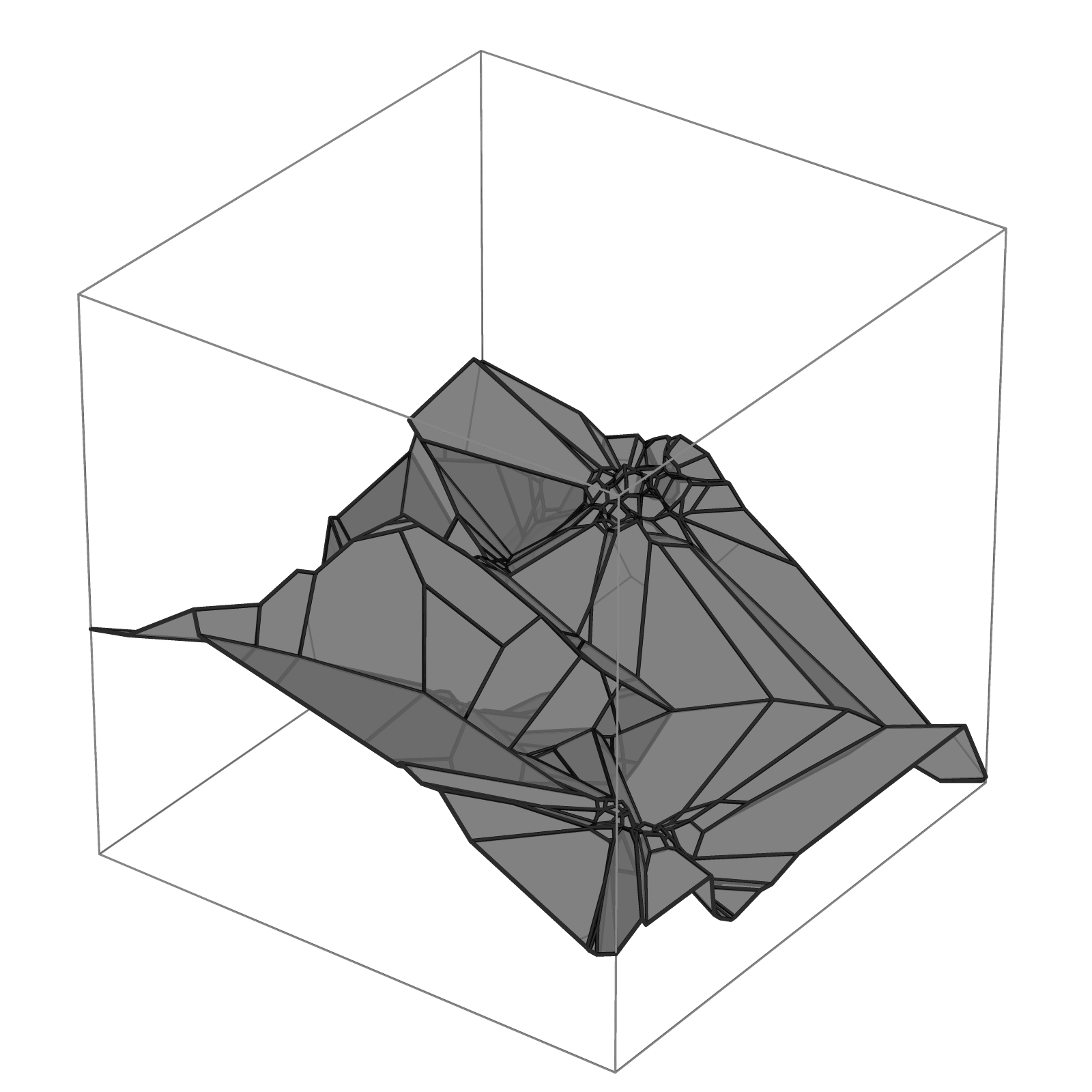}}\hfil
	\subfloat[$\lambda=100$]{\includegraphics[width=0.24\textwidth]{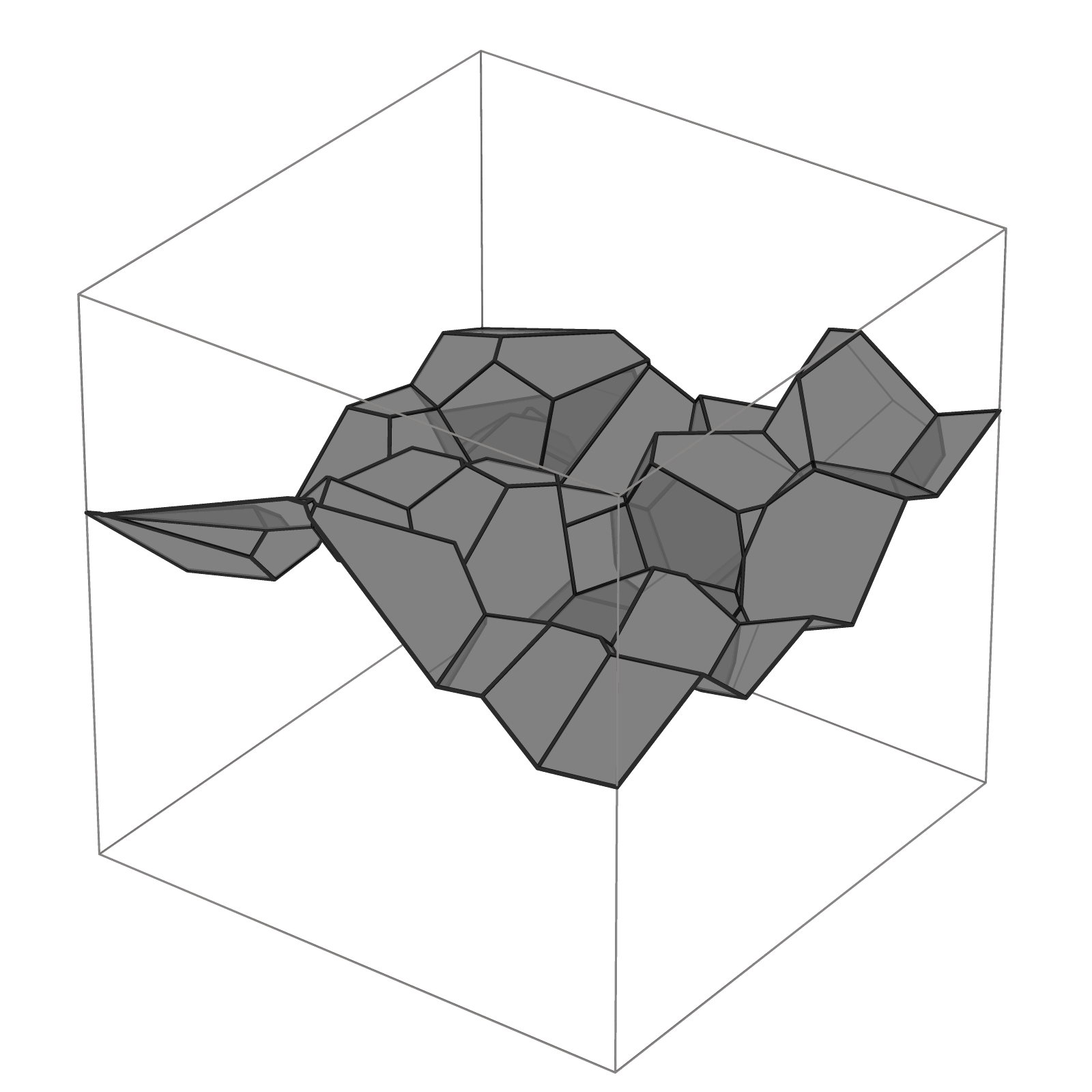}}\hfil
	
	\subfloat[$\lambda=1000$]{\includegraphics[width=0.24\textwidth]{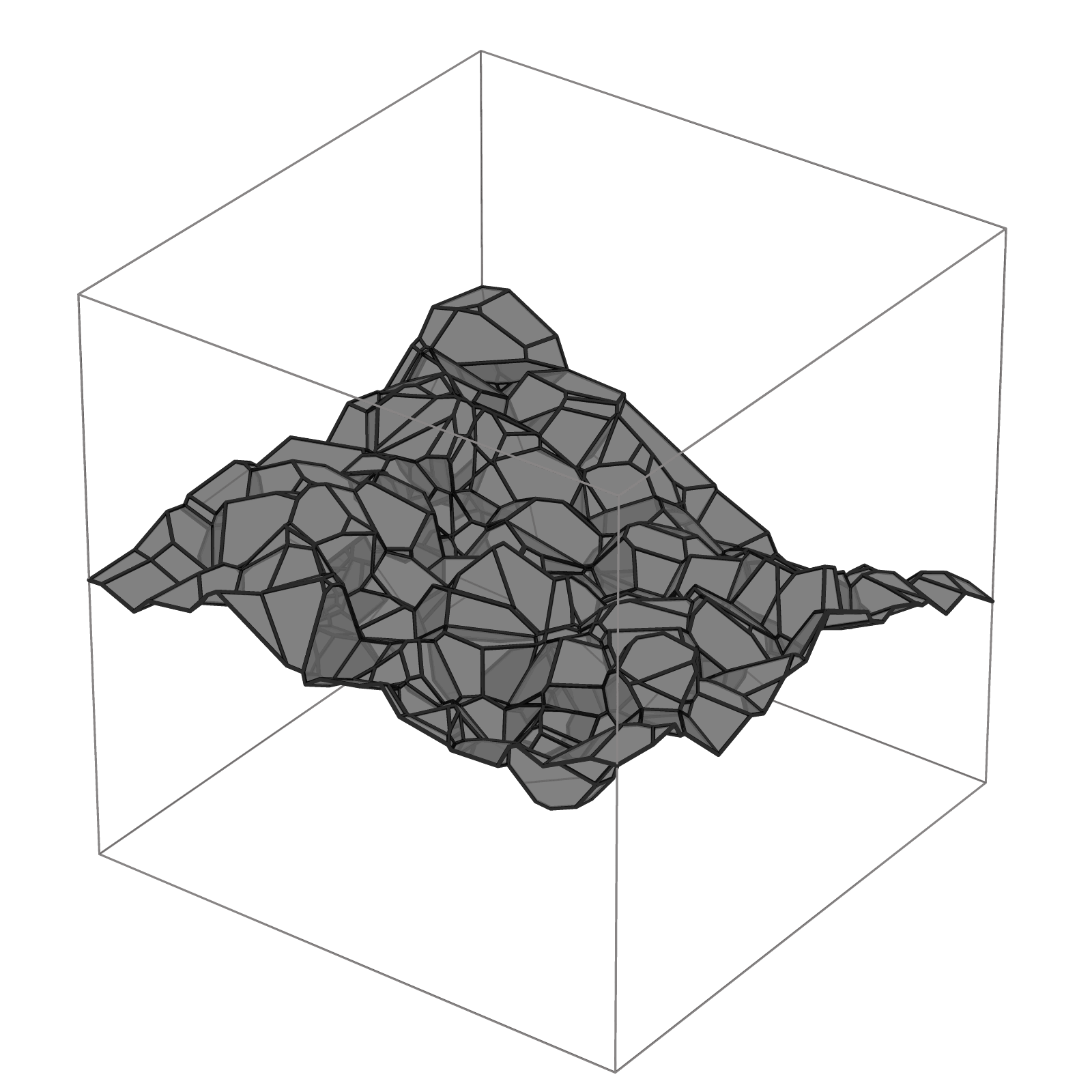}}\hfil
	\subfloat[$\lambda=20,\mu=50,r=0.1$]{\includegraphics[width=0.24\textwidth]{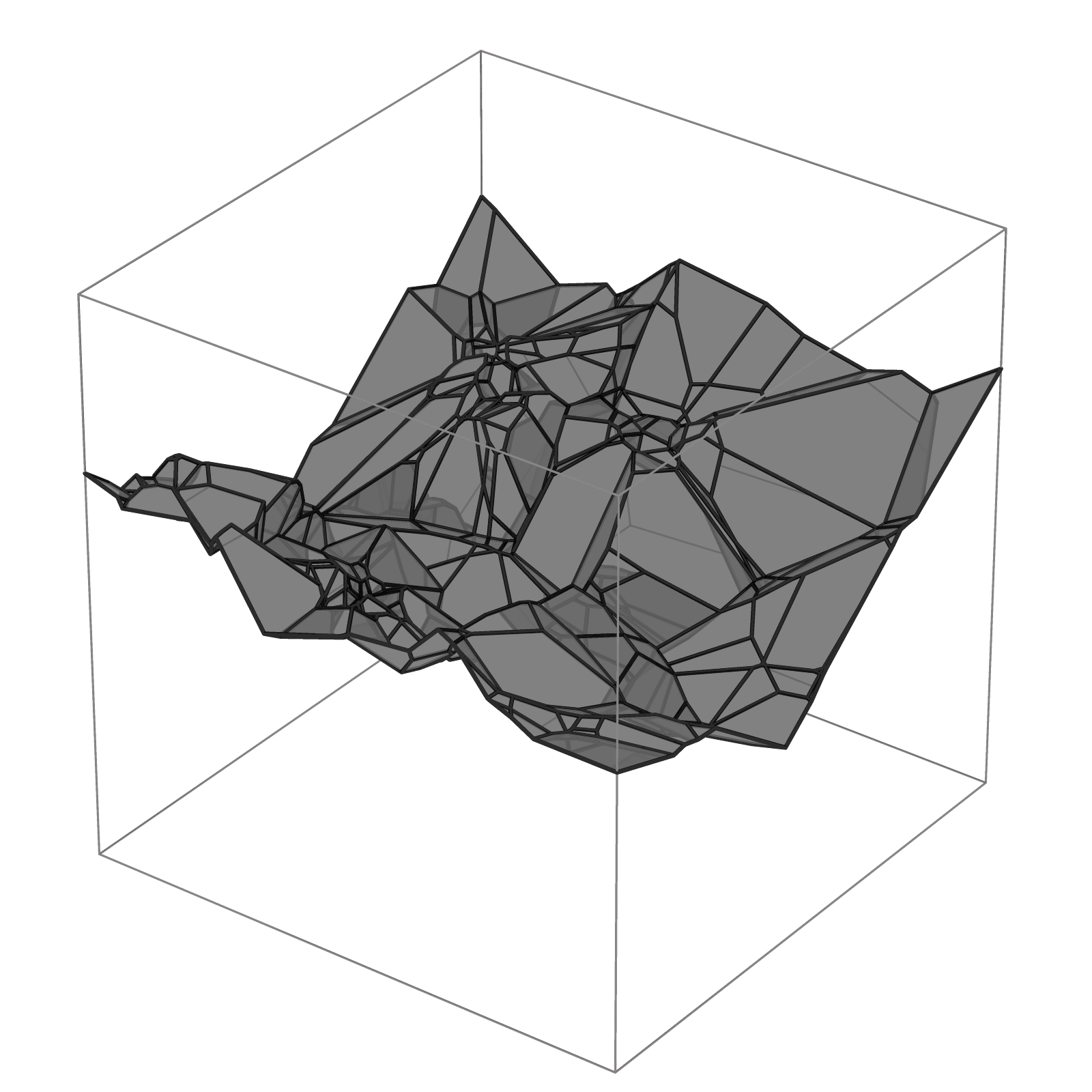}}\hfil
	\subfloat[$\lambda=1000$]{\includegraphics[width=0.24\textwidth]{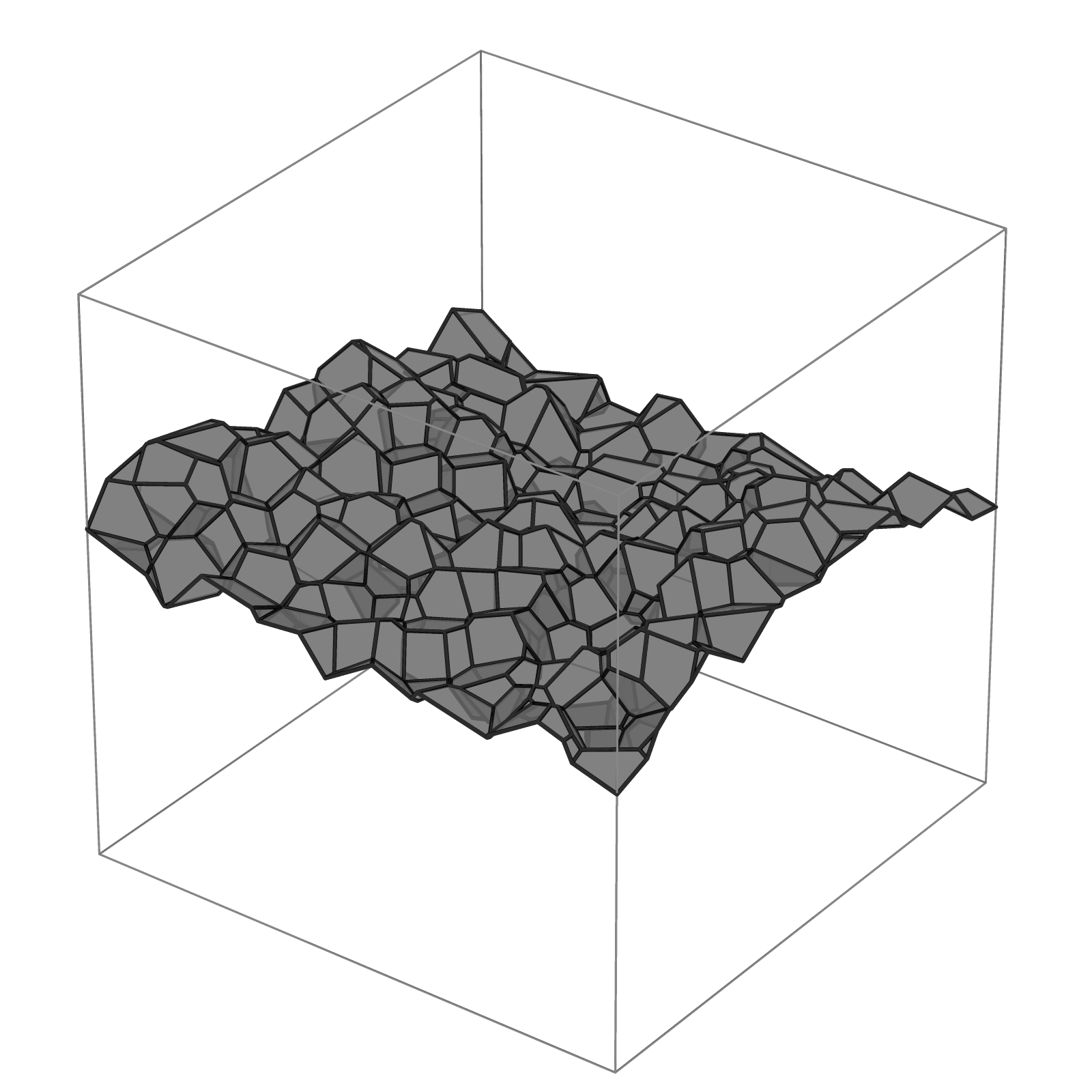}}\hfil
	
	\subfloat[$\lambda=5000$]{\includegraphics[width=0.24\textwidth]{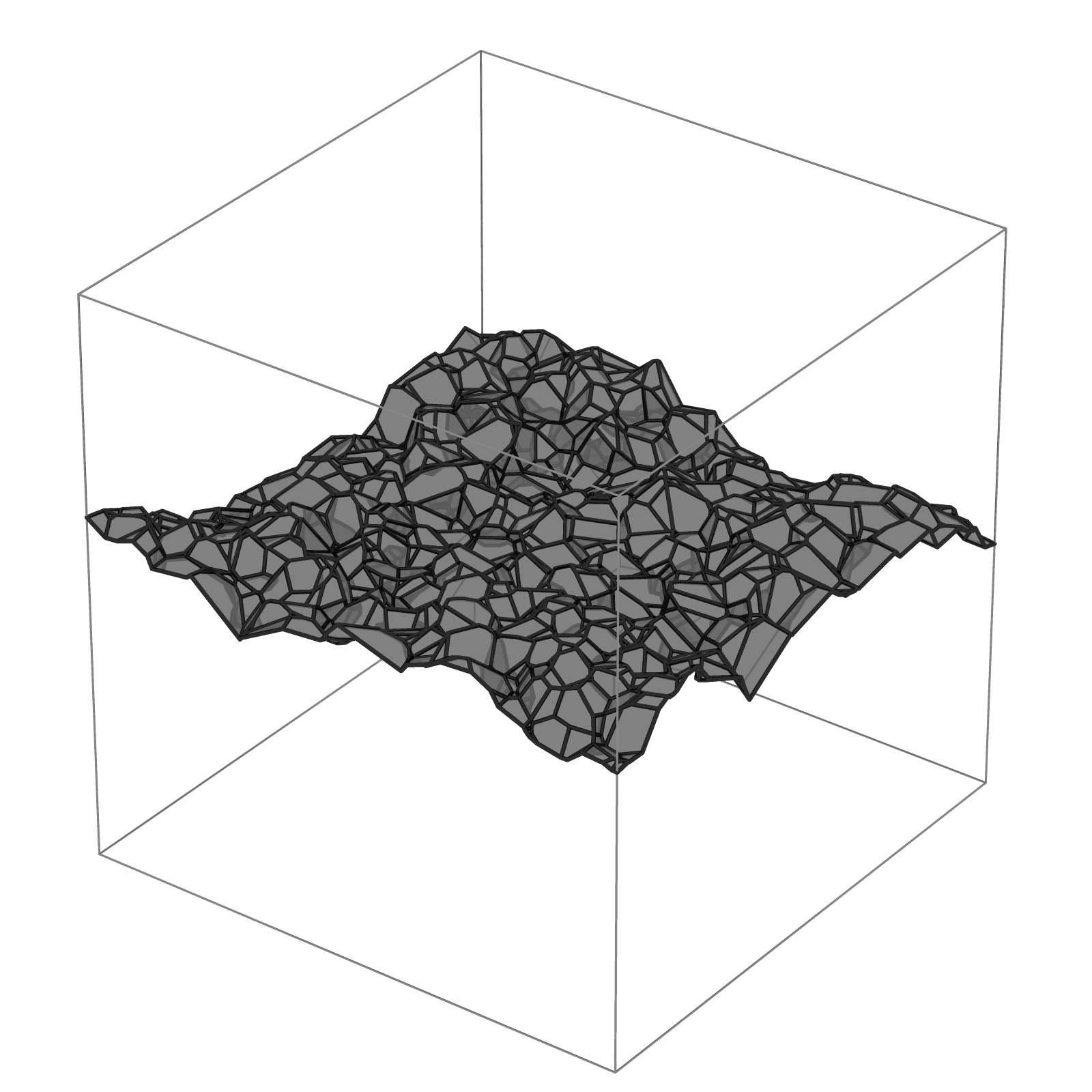}}\hfil
	\subfloat[$\lambda=50,\mu=100,r=0.1$]{\includegraphics[width=0.24\textwidth]{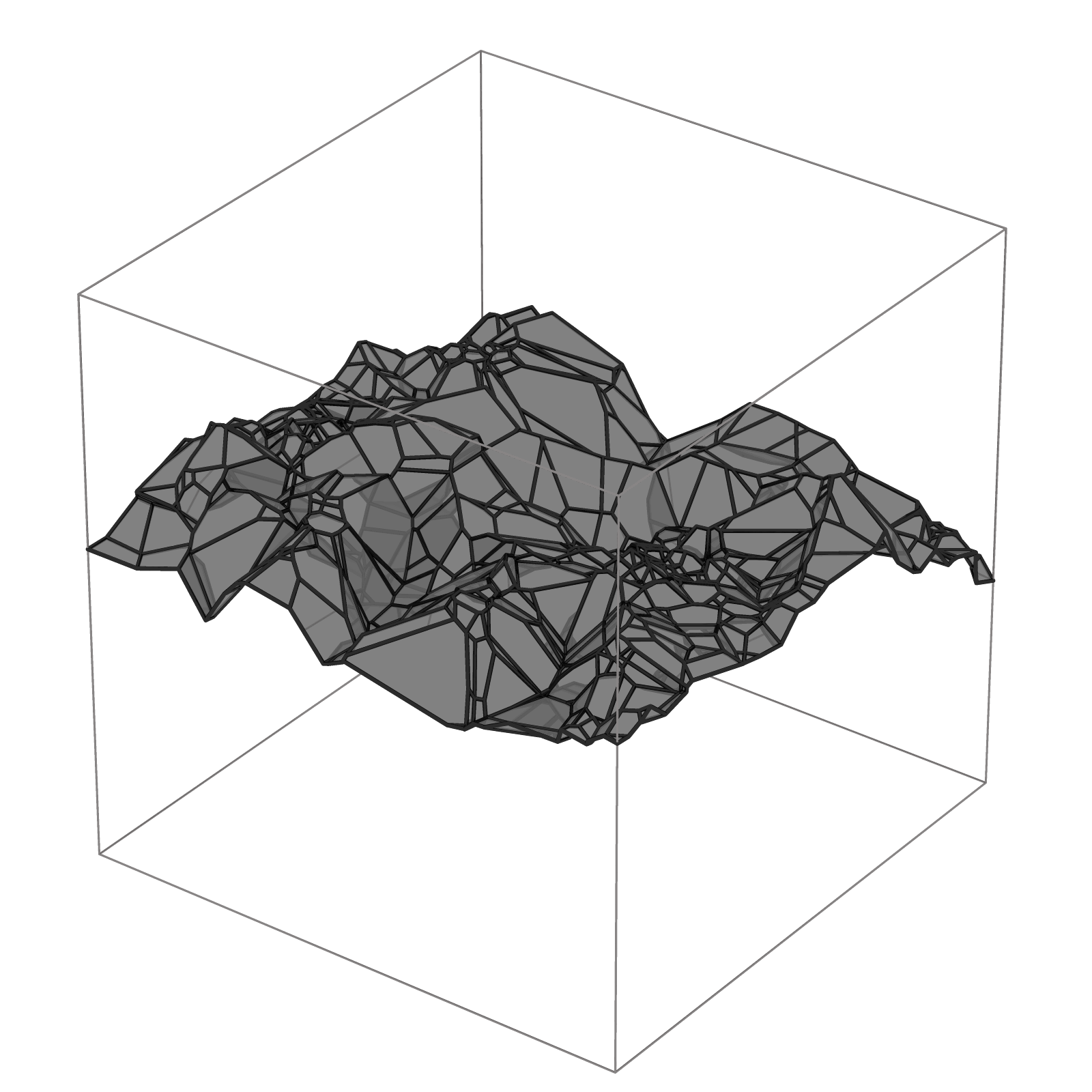}}\hfil
	\subfloat[$\lambda=5000$]{\includegraphics[width=0.24\textwidth]{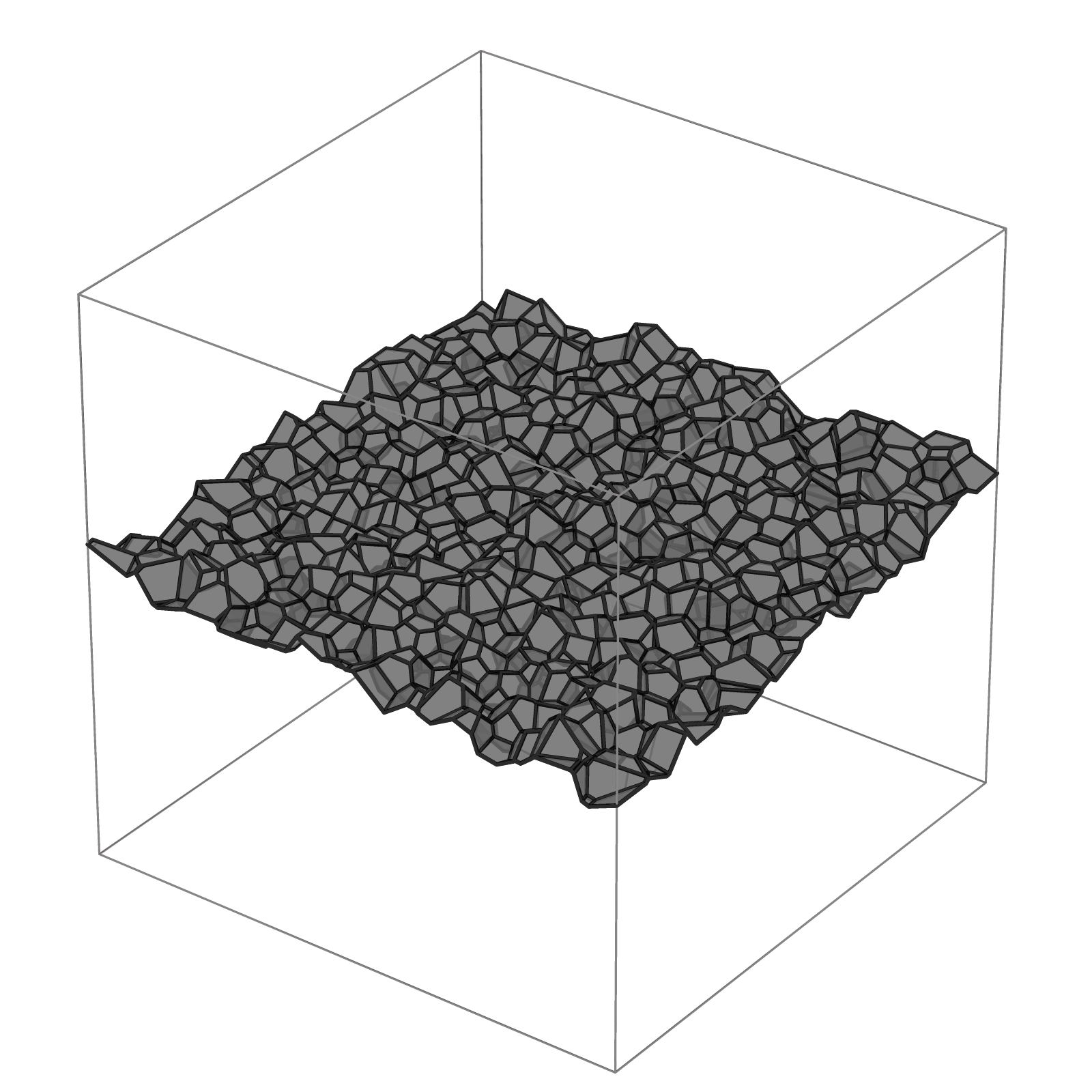}}\hfil
	
	\caption{Minimum-weight surfaces in Voronoi diagrams bounded by $Q=[0,1]\times[0,1]\times[0,1]$. The generators are realizations of Poisson point processes (first column) with intensity $\lambda$, Matern cluster processes (second column) with parent and daughter point process intensities $\lambda$ and $\mu$, respectively, and cluster radius $r$, and hardcore processes (third column) with constant radii, intensity $\lambda$ and $60\%$ volume fraction.}\label{fig:minsurfPP}
\end{figure*}

Moreover, Fig. \ref{fig:minsurfPP} shows minimum-weight surfaces in Voronoi diagrams generated by Poisson point processes, Mat\'ern cluster processes and regular processes obtained by a force-biased sphere packing \cite{forcebiased1,forcebiased2} with a volume fraction of $60\%$. Arcs and facets were weighted by their lengths and areas, respectively. For the cluster process, we observe a bimodal distribution of the facet areas. The surfaces resulting from the regular model are far more homogeneous in facet size and shape than those obtained from Poisson point processes.


For the Voronoi diagram generation we use the C++ library voro++\cite{voropp} and for solving the integer program we use the C++ library GLPK \cite{glpk}. 

The runtime for computing a minimum-weight surface depends mainly on the size of the arc-facet incidence matrix, that is, on the number of facets in the cell complex. For a Poisson Voronoi diagram in $[0,1]\times[0,1]\times[0,1]$ with intensity 500 the expected runtime is around 12 seconds. It is obtained on a Red Hat Enterprise Linux Workstation 7.9 with an Intel(R) Xeon(R) CPU E5-2680 v2 2.8GHz (10 cores) and 125 gigabytes of RAM.

\subsection{Discretization}\label{disc}

The method described in Section \ref{crackgen} outputs a set of vertices of convex facets. In this section, we describe a method to transfer this representation to a discrete image.

Let $I$ denote a 3d label image and $J$ a 3d binary image, both of size $d_1\times d_2\times d_3$. The discretization procedure is given as follows.

\begin{enumerate}
	\item Compute a minimum-weight Voronoi surface in a cuboid of size $d_1\times d_2\times d_3$.
	\item Discretize the Voronoi diagram. For every voxel $(p,q,r)$ do: Set $I(p,q,r)=l$ if $(p,q,r)$ is contained in cell $l$.
	\item Discretize the minimum-weight surface. For every two neighboring voxels $(p,q,r),$ $(p',q',r')$ (with respect to the 26-neighborhood) do: Set $J(p,q,r)=1$ if $I(p,q,r)=j$ and $I(p',q',r')=k$ for generators $j,k$ whose cells share a facet that is part of the minimum-weight surface. Output $J$.
\end{enumerate}

\begin{figure}[h]
	\centering
	\subfloat{\includegraphics[width=0.3\textwidth]{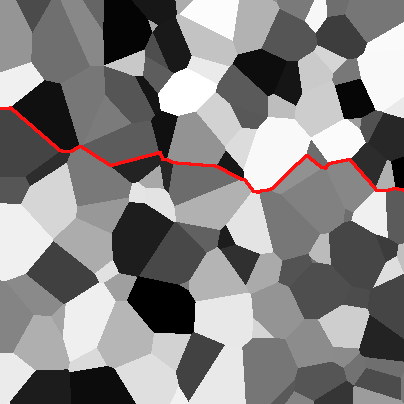}}\quad
	\subfloat{\includegraphics[width=0.3\textwidth]{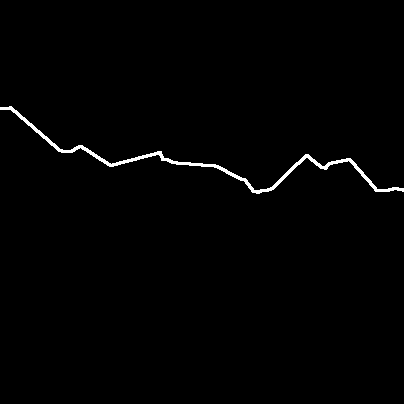}}
	\caption{Slice view visualizing the discretization procedure. Left: Label image obtained from discretizing the Voronoi diagram. Red voxels indicate facets that are part of the minimum-weight surface. Right: Output binary image.}\label{fig:discret}
\end{figure}

The procedure is visualized in Fig. \ref{fig:discret}. It yields a binary image whose foreground is a piecewise planar structure of constant width. However, real crack structures usually are far more complex as has been pointed out in Section \ref{sec:crackModeling}. In the following, we propose techniques to account for these observations.


\subsubsection{Adaptive dilation}

\begin{figure}[h]
	\centering
	
	\subfloat{\includegraphics[width=0.3\textwidth]{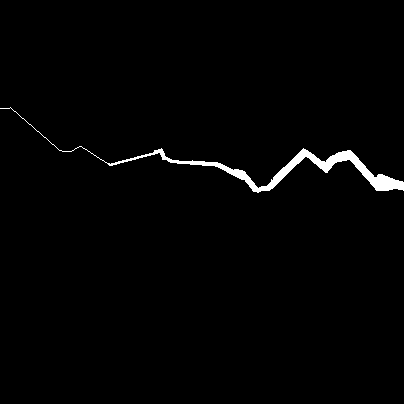}}\quad
	\subfloat{\includegraphics[width=0.3\textwidth]{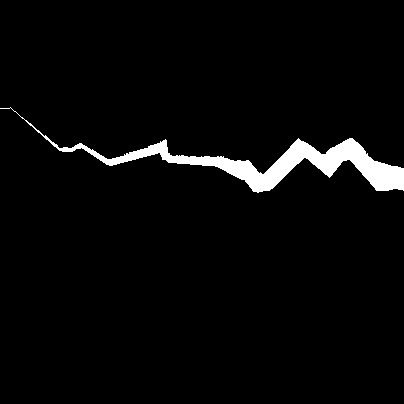}}\quad
	\caption{Adaptive dilation applied to the input image with $p=0.02$ (left) and $p=0.05$ (right).}\label{fig:endvertadaptDil}
\end{figure}

Our goal is to model cracks of varying thickness. We propose a procedure to dilate the foreground of image $J$: Every $x$-slice of $J$ is dilated separately and iteratively. We choose a quadratic structuring element of size $2\times 2$. The number of iterations depends on a random walk with Bernoulli-distributed increments and index set $\{0,1,\ldots, d_1\}$. The increments are either 1 with probability $p$ or 0 with probability $1-p$. Thus, the crack thickness can be controlled via parameter $p$. The procedure is visualized in Fig. \ref{fig:endvertadaptDil} for different choices of $p$.
Note that the random walk can be substituted by any suitable stochastic process, for example to produce decreasing crack widths.

\subsubsection{Microstructure modeling}\label{micromodel}

\begin{figure}[h]
	\centering
	\subfloat{\includegraphics[width=0.3\textwidth]{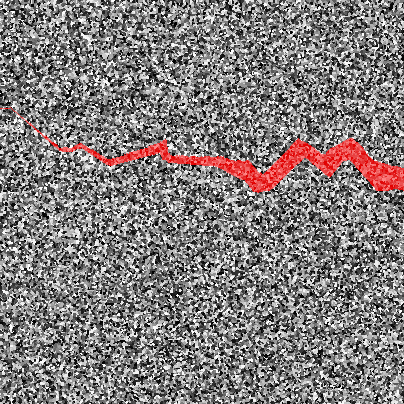}}\quad
	\subfloat{\includegraphics[width=0.3\textwidth]{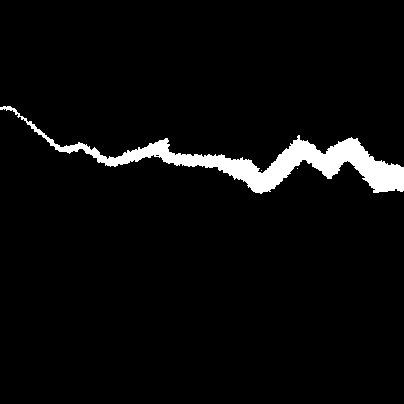}}\quad
	\caption{Microstructure generation. Left: Discretized Poisson-Voronoi diagram and input crack (red). Right:  Set of Voronoi cells (white) that are intersected by the input crack.}\label{fig:discretmicro}
\end{figure}

In order to model the rough microstructure on the boundary of cracks, we compute a second (Poisson-) Voronoi diagram with a higher intensity than the one used for the computation of the minimum-weight surface. Then, for every foreground voxel in the dilated crack image $J$, we identify the Voronoi cell it is contained in. The whole cell is then discretized with voxel value 1 according to the approach described in Section \ref{disc}. The procedure is visualized in Fig. \ref{fig:discretmicro}. Afterwards, we apply a median filter to the resulting image. 

\subsubsection{Crack branching}

Crack branches emerge when a crack splits into two or more cracks. Often, the thickness of these branches lies in a range of 1-2 voxels. 

Branching cracks can be modeled by combining two minimum-weight surfaces obtained from different cycles on the cuboid. If the underlying set of generators for the Voronoi diagram is identical, the surfaces may share several facets, see Fig. \ref{fig:discretmicro2}.

\subsubsection{Crack embedding}

\begin{figure}[h]
	\centering
	\subfloat{\includegraphics[width=0.3\textwidth]{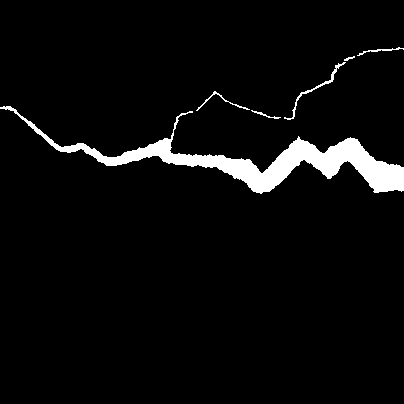}}\quad
	\subfloat{\includegraphics[width=0.3\textwidth]{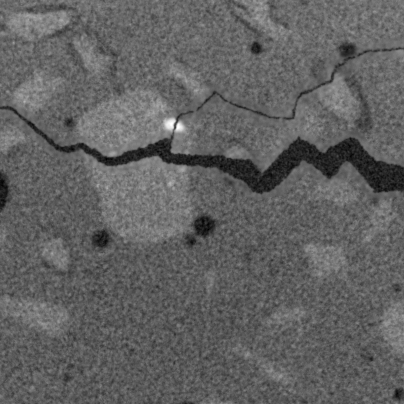}}\quad
	\caption{Crack embedding. Left: Ground truth image after applying a median filter, right: synthesized crack image.}\label{fig:discretmicro2}
\end{figure}

\begin{figure}[h]
	\centering
	\subfloat{\includegraphics[width=0.4\textwidth]{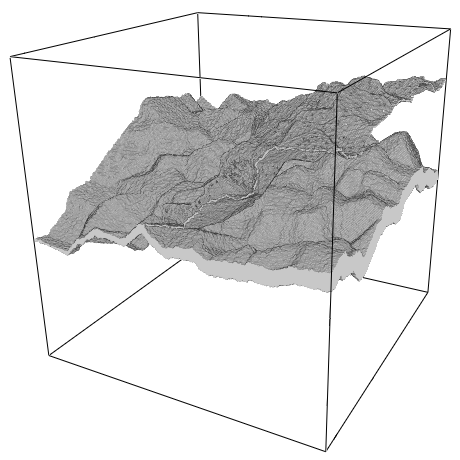}}
	\caption{3d volume rendering of the simulated crack structure.}\label{fig:discretmicro3}
\end{figure}

An approach for embedding crack structures in real 3d concrete images has been proposed in \cite{ourPaper}. We extract image patches of the same size as the ground truth images from the real CT images. These patches are multiplied voxelwise with the inverse ground truth images. This leads to crack voxels having grayvalue 0 while the background does not change. Cracks and air pores both consist of air. Therefore, they should possess the same grayvalue distribution. We assume these grayvalues to be i.i.d. normally distributed. Mean and standard deviation are estimated via sample mean and sample standard deviation of the empirical distribution of air pore grayvalues. Then the crack voxels are simulated according to that distribution. To smooth the transition between background and crack, we apply a Gaussian filter to crack voxels and all voxels in their 26-neighborhood. The final image together with its ground truth is given in Fig. \ref{fig:discretmicro2}. A corresponding 3d rendering is given in Fig. \ref{fig:discretmicro3}.

\section{Conclusion}

In this work, we have presented a novel method to generate artificial crack images. It includes the generation of a macrostructure via minimum-weight surfaces and a discretization procedure for generating its microstructure.

The shape and size of the output can be controlled by several parameters. Thus it allows for the generation of a wide range of surface structures.

Our next steps include the generation of a full semi-synthetic data set for training machine learning models and evaluating segmentation methods on multiscale crack images. 

Furthermore, the model may be extended to include the grayvalue information of real concrete images. We can assume that cracks, when propagating through concrete, take the path of least resistance. Certain parts of the concrete mixture are less prone to cracking than other parts. In particular, this holds true for parts with a higher density. Therefore, facet weights may be derived from the mean voxel grayvalue in their vicinity.\\


\textbf{Acknowledgements:} This work was supported by the German Federal Ministry of Education and Research (BMBF) [grant number 05M2020 (DAnoBi)].

The authors would like to thank Prof. Dr. Sven O. Krumke for his helpful suggestions.

\bibliographystyle{plain}
\bibliography{lit}

\end{document}